\documentstyle[11pt]{article}

\newcommand{\bd}{\begin{document}}
\newcommand{\ed}{\end{document}}
\newcommand{\bc}{\begin{center}}
\newcommand{\ec}{\end{center}}
\newcommand{\vs}{\vspace}
\newcommand{\hs}{\hspace}
\newcommand{\bq}{\begin{quote}}
\newcommand{\eq}{\end{quote}}
\newcommand{\lb}{\linebreak}
\renewcommand{\pb}{\pagebreak}
\newcommand{\beq}{\begin{equation}}
\newcommand{\eeq}{\end{equation}}
\bd
\addtolength {\oddsidemargin} {-2.5cm}
\setlength{\topmargin}{-1.3cm}
\setlength{\parskip}{1pc}
\setlength{\parindent}{0pc}

\rm
 
\title 
{{\bf Frustrated Synchronization in Competing Drive-Response
Coupled Chaotic Systems}}

\vs{1cm}
\author{Sitabhra Sinha\\}

\date{Machine Intelligence Unit, Indian Statistical Institute\\
203 B. T. Road, Calcutta-700035, INDIA\\
E-mail: res9425@isical.ac.in}

\maketitle
{\abstract
Chaotic systems can be synchronized by linking them to a common signal,
subject to certain conditions. However, the presence of multiple driving
signals coming from different systems, give rise to novel behavior.
The particular case of Lorenz systems, with two independent systems 
driving another system through drive-response coupling has been studied
in this paper. This is the simplest arrangement
which shows the effect of ``frustrated synchronization'' due to
competition between the two driver systems. The resulting
response system attractor deviates significantly from
the conventional Lorenz attractor. A new measure of desynchronization is
proposed, which shows a power-law scaling relation with the competition 
parameter.}\\

PACS no.: 05.45.+b.\\

Keywords: Chaos, Synchronization, Frustration, Competition, Lorenz System.
\special{I+}
\pb

\section{Introduction}

The synchronization of chaotic systems is a difficult problem
owing to their extremely sensitive dependence on initial conditions.
Any initial correlation present between identical systems,
starting from very close initial conditions, exponentially decrease to
zero with time. Thus, for all practical purposes, any initial synchronization
between the systems is bound to disappear rapidly. 
In recent times, however, some methods of achieving synchronized behavior
between chaotic systems have been proposed. Pioneering work in this respect
has been done by Pecora and Carroll \cite{Pecora90}, who used
the concept of a response system
locking on to a driver system. So far, such studies have been limited to
driving a response system by a single driver system. However, the knowledge
gained from studying such simple systems may not be adequate to give us an
idea as to how systems consisting of multiple independent driver systems,
competing with each other to synchronize the same response system,
will behave. The Pecora-Carroll driving mechanism
can be seen as the ``strong-coupling''
limit of a general scheme of directionally- oriented couplings in
a network of chaotic elements.

The synchronization of bidirectionally coupled chaotic systems is
stable provided the coupling strength is at least half
the Lyapunov exponent of the system \cite{Fujisaka}.
One-way coupling (or, driving one chaotic system by another)
can also lead to synchronization, provided
certain conditions are satisfied \cite{Pecora90}, \cite{
Pecora91}, \cite{HeVaidya}.
The drive-response method consist of the following steps.
First an $n$-dimensional autonomous system
$$
{\frac {d{\bf x}}{dt}}~=~{\bf F(x)},
$$
is divided into two parts, driving (${{\bf x}_d}$) and 
responding (${{\bf x}_r}$):
$$
{\frac {d{{\bf x}_d}}{dt}}~=~{\bf g}({\bf x}_{d},{\bf x}_{r}) ,~~~
{\frac {d{{\bf x}_r}}{dt}}~=~{\bf h}({\bf x}_{d},{\bf x}_{r}) ,
$$
where, ${\bf x}_d ~=~ (x_1, {\ldots}, x_m)$, ${\bf g}~=~ [ f_1 ({\bf x}), {\ldots},
f_m ({\bf x}) ]$, ${\bf x}_r ~=~(x_{m+1}, {\ldots}, x_n)$ and ${\bf h}~=~ 
[ f_{m+1} ({\bf x}), {\ldots}, f_n ({\bf x}) ]$. A replica subsystem
${\bf x}^{\prime}_{r}$ identical to ${\bf x}_{r}$ is then created and driven
with the ${\bf x}_{d}$ variables of the original system.
Therefore, the replica subsystem equations are,
$$
{\frac {d{\bf x}^{\prime}_{r}}{dt}}~=~{\bf h}({\bf x}_d,~{\bf 
x}^{\prime}_{r}).
$$
The responding subsystems ${{\bf x}_r}$ and ${\bf x}^{\prime}_{r}$ will
synchronize only if $\delta {{\bf x}_r} = |{{\bf x}_r} -
{\bf x}^{\prime}_{r}| \rightarrow 0$. According to Pecora and Carroll,
this occurs if and only if the conditional Lyapunov exponents of the
${\bf x}_r$ subsystem are all negative. 

Drive-response synchronization has been realized in various electrical
circuit experiments.  It has also been used in experiments of secure
communication where a chaotic masking signal is added to the
transmitted signal. It is then recovered at the receiving end by
subtracting the chaotic signal regenerated by synchronization \cite{Cuomo}.

Besides the Pecora-Carroll method, other synchronization procedures
have also been proposed.
Of these, the Variable Control Feedback (VCF) method is of particular
interest, as it can be used for both control and synchronization of chaos
\cite{HuQuHe}. In fact, the Pecora-Carroll method turns out to be a
special limiting case of this method.
VCF consists of adding a feedback term to a dynamical system
to guide it into some prescribed state. If ${\frac{d{\bf{\rm x}}}{dt}}
= {\bf {\rm F(x)}}$ be an $n$-dimensional 
dynamical system and ${\bf {\rm x^{*}}}$ be the desired state 
to which the system has to be brought, then
VCF involves modifying the system dynamics to:
$$ {\frac{d{\bf{\rm x}}}{dt}} = {\bf {\rm F(x)}} - {\lambda} ({\bf {\rm x - x^{*}}}) $$
where ${\lambda}$ is the set of $n$ feedback multipliers. If
${\bf {\rm x^{*}}}$ be the output of a chaotic system 
${\bf {\rm F^{\prime}(x)}}$, then the system synchronizes with
${\bf {\rm F(x)}}$. In the large-${\lambda}$ limit, VCF reduces to the
Pecora-Carroll method. Specifically, the feedback parameters for the
driving subsystem variables, ${\lambda_d} \rightarrow \infty$, while the
remaining $ \lambda $s are set to zero.

In this paper some observations have been reported on the
attractor structure of a chaotic system which has been
subjected to simultaneous synchronization by two other identical chaotic
systems competing with each other. Section 2 introduces the model used
for studying competition among synchronizing chaotic systems and
includes a short analysis of the fixed points and their stability.
Section 3 contains the results of computer simulations of the system.
Finally, possible directions of future research and the 
relevance of this type of research to the theory of neural computation
are discussed.
 
\section{Competition among synchronizing Lorenz systems}
The investigation of competition among synchronizing chaotic systems was
carried out using the Lorenz system of
equations \cite{Lorenz}, \cite{Sparrow}. This
well-known paradigm of chaos is defined by the following set of equations:
\beq
\frac{dx}{dt} ~=~ \sigma ~(y ~-~ x),
\eeq
\beq
\frac{dy}{dt} ~=~ rx ~-~ y ~-~ xz,
\eeq
\beq
\frac{dz}{dt} ~=~ xy ~-~ bz,
\eeq
where, $\sigma$, $r$ and $b$ are real, positive parameters.
There are three fixed points for this system: ${\rm F}_1 = (0, 0, 0)$, 
${\rm F}_2 = ({\sqrt{b(r-1)}}, {\sqrt{b(r-1)}},~r-1)$, and ${\rm F}_3 = 
(~-{\sqrt{b(r-1)}}, -{\sqrt{b(r-1)}},~r-1)$.
The local stability of the fixed point ($x_f~,~y_f~,~z_f$) is  determined by
the eigenvalues of the Jacobian
\beq
{\bf{\rm J}} ~=~  \left| \begin{array}{ccc}
         -~{\sigma}  & {\sigma} & 0\\
	 (~r~-~z_f~)   & -~1      & -~x_f\\
          y_f          & x_f        & -~b
         \end{array} \right|.
\eeq
Evaluation of the matrix shows that for $0 < r < 1$, ${\rm F}_1$ is the 
only stable fixed point. For $r > 1$, ${\rm F_1}$ becomes unstable and the
phase-space trajectory of the system converges to either ${\rm F}_2$
or ${\rm F}_3$. For $r > r_c = \sigma (\sigma + b + 3)/(\sigma - b - 1) $
the system's trajectory perpetually wanders along the extremely complicated
structure of the stable and unstable manifolds of the fixed points,
exhibiting chaotic behavior.

For the present work the effect of two driving systems, designated as driving
systems 1 ($x_1, y_1, z_1$) and 2 ($x_2, y_2, z_2$),
competing to
synchronize a responding system ($x_3, y_3, z_3$) was studied.
The responding system was driven using the $y$ variable. A competition
parameter $a$ was defined to indicate the strength of the driving
systems relative to each other. The maximum value of $a$ was normalized
to unity. Therefore, the $y$ variable of the responding system was
defined in terms of the two driving systems as:
\beq
y_3 = a y_1 + (1 - a) y_2.
\eeq
We consider first the case where the two driving systems have
the same $r$-parameter value, and then, the more general
case, where the two $r$-values are different ($r_1$ and $r_2$, say).
The $\sigma$ and $b$-parameter values are considered to be the same in
all cases.

{\bf Case I:} $r_1 = r_2 = r $

It is obvious that for $a = 1$ the responding system synchronizes with
driver system 1, whereas for $a = 0$, it synchronizes with system 2.
The attractor of the response system, is identical to that of the
conventional Lorenz system (fig. 1(a)).
For $0< a< 1$, the responding system ($x_3,y_3,z_3$) has nine fixed points:\\
${\bf {\rm F}}_1 ~=~ (0, 0, 0)$,\\
${\bf {\rm F}}_2 ~=~ ({\sqrt{b(r-1)}}, {\sqrt{b(r-1)}}, r-1)$,\\ 
${\bf {\rm F}}_3~=~( -{\sqrt{b(r-1)}}, -{\sqrt{b (r-1)}}, r-1)$,\\ 
${\bf {\rm F}}_4$~=~($a {\sqrt {b(r-1)}}$,$ a {\sqrt {b (r-
1)}}$, $a^2 (r-1) )$,\\
${\bf {\rm F}}_5~=~((1 - a) {\sqrt {b(r-1)}}, (1 - a) {\sqrt {b(r-1)}}, 
(1 - a)^2 (r-1) )$,\\
${\bf {\rm F}}_6~=~( -a {\sqrt {b(r-1)}}, -a {\sqrt {
b(r-1)}}, a^2 (r-1) )$,\\
${\bf {\rm F}}_7~=~(-(1 - a) {\sqrt {b(r-1)}}, -(1 - a) {\sqrt {b(r-1)}}, 
(1 - a)^2 (r-1) )$,\\
${\bf {\rm F}}_8~=~( (2a - 1) {\sqrt{b(r-1)}}, (2a - 1) {\sqrt 
{b(r-1)}}, (2a - 1)^2 (r - 1) )$,\\
${\bf {\rm F}}_9~=~
( -(2a - 1) {\sqrt{b(r-1)}}, -(2a -
1) {\sqrt {b(r-1)}}, (2a - 1)^2 (r - 1) )$.

Note that the first three fixed points are those of the uncoupled Lorenz
system. To find out about the stability of these fixed points we need to
calculate the eigenvalues of the corresponding Jacobian, ${\bf {\rm 
J^{\prime}}}$. The partially block-diagonal form of the matrix makes the
calculation easy:
\beq
{\bf{\rm J^{\prime}}} ~=~  \left| \begin{array}{ccc}
              {\bf {\rm J}} & 0_{3 \times 3} & 0_{3 \times 2} \\
              0_{3 \times 3} & {\bf {\rm J}} & 0_{3 \times 2} \\
              {\bf {\rm A}} & {\bf {\rm B}} & {\bf{\rm J}}_R 
              \end{array} \right|,
\eeq
where, ${\bf {\rm J}}$ is the Jacobian (eqn. 4) of the unperturbed 
Lorenz system of equations, $0_{m \times n}$ is a null 
matrix having $m$ rows and $n$
columns, and the other matrices are defined as,
\beq
{\bf {\rm A}} ~=~ \left| \begin{array}{ccc}
                    0 & a~{\sigma} & 0 \\
		    0 & a~x_{f_3}    & 0
                  \end{array} \right|,
\eeq
\beq
{\bf {\rm B}} ~=~ \left| \begin{array}{ccc}
                  0 & (~1~-~a~)~{\sigma} & 0\\
                  0 & (~1~-~a~)~x_{f_3} & 0 
                  \end{array} \right|,
\eeq		
and,
\beq
{\bf {\rm J}}_R ~=~ \left| \begin{array}{cc}	     
                -~{\sigma} & 0 \\
	        a~y_{f_1}~+~(~1~-~a~)~y_{f_2} & -~b
                \end{array} \right|.
\eeq
Here $f_k$ refers to the fixed point of the $k$th Lorenz system.

For $0 < r < 1$, the only stable fixed point is ${\rm F_1}$.
For $r > 1$, ${\rm F_1}$ loses its stability,
and there are four new stable fixed points: 
${\rm F_2, F_3, F_8}$  and ${\rm F_9}$.
For $r > r_c = \sigma (\sigma + b + 3)/(\sigma - b - 1) $,
these fixed points lose their stability and the system
shows only chaotic behavior.
The most interesting instance is that of $a=0.5$,
where maximal competition occurs.
In this case, ${\rm F_8 = F_9 = F_1}$, ${\rm F_4 = F_5}$
and ${\rm F_6 =F_7}$ (fig. 2).
The attractor of the responding system
is found to be stretched over its 3-dimensional phase space showing
an extremely tangled structure (fig. 1(b)). This is due to the
extremely complicated motion of the response system trajectory along the
stable and unstable manifolds of the fixed points ${\rm 
F_1, F_2, F_3, F_4 ~and~ F_6}$. The coupling with 
driver system 1 tries to force
the response system into synchronization with it, but at the same time,
the coupling with driver system 2 desynchronizes the trajectory. The
synchronization is therefore `frustrated' by the competition between the
two driver systems. The ``frustrated'' response system attractor
reduces to the conventional Lorenz attractor if $a \rightarrow 
0 ~{\rm or}~ 1$, when competition is absent. 

The attractor structure is found to be quite robust.
If we start from two different initial conditions for the responding system,
$(x,y,z)$ and $(x^{\prime},y^{\prime},z^{\prime})$, say, then for stable
synchronization, the two respective trajectories should converge rapidly.
However, whereas in the Pecora-Carroll case, convergence occurs to the
standard Lorenz attractor, in this case, both the trajectories
converge to the ``frustrated''attractor. 

The stability of synchronization can be demonstrated analytically
by linear stability analysis of the error dynamics. Defining the dynamical
error between two response system trajectories (${\bf {\rm x}}$ and
$ {\bf {\rm x^{\prime}}}$) which have different initial conditions, as
${\bf {\rm e = x - x^{\prime} }}$, the error equations can be written as:
\beq
{\frac {d e_x}{dt}} = - \sigma e_x,
\eeq
\beq
e_y = 0,
\eeq
\beq
{\frac {d e_z}{dt}} = ( a y_1 + (1-a) y_2 ) e_x - b e_z.
\eeq
Here we have assumed that the equation parameters for the two systems are
identical. The error system of equations has an equilibrium point at
${\bf {\rm e}}$ = (0, 0, 0), which corresponds to perfect synchronization.
The local stability of synchronization can then be checked by looking at
the eigenvalues of the Jacobian of the error equations:
\beq
{\bf {\rm J}}_R ~=~ \left| \begin{array}{cc}	     
                -~{\sigma} & 0 \\
	        a~y_1~+~(~1~-~a~)~y_2 & -~b
                \end{array} \right|
\eeq
The eigenvalues are $ - \sigma$ and $ - b $, which are the conditional Lyapunov
exponents of the response system. As both eigenvalues are negative, the
synchronization is locally stable, and any difference in
initial conditions rapidly goes to zero.
Note that, this does not prove the global stability of the
synchronized state. However, simulations have verified that even in the
presence of large deviations in initial conditions, synchronization
with the ``frustrated'' trajectory is achieved. This indicates that,
although exact synchronization with the driver system cannot be achieved,
the ``frustrated'' system can still be used for secure
communication through chaotic masking. This has been established through
simulations reported below.

{\bf Case II:} $r_1 \neq r_2 $

When the value of the $r$-parameter of the two driving systems is not the
same, the fixed points are given by:\\
${\bf {\rm F_1}} = (0, 0, 0),\\
{\bf {\rm F_2}} = (a {\sqrt{b(r_1-1)}} + (1 - a) {\sqrt{b(r_2-1)}}, 
a {\sqrt{b(r-1)}} + (1 - a) {\sqrt{b(r_2-1)}}, a^2 (r_1 - 1) + (1-a)^2 (r_2-1)
+ 2a (1 - a) {\sqrt (r_1-1) (r_2-1)}),\\
{\bf {\rm F_3}} = (-a {\sqrt{b(r_1-1)}}-
(1-a) {\sqrt{b(r_2-1)}}, -a {\sqrt{b(r-1)}} - (1-a) {\sqrt{b(r_2-1)}}, a^2 
(r_1 - 1) + (1-a)^2 (r_2 - 1) + 2a (1 - a) {\sqrt (r_1-1)(r_2-1)}),\\
{\bf {\rm F_4}} = (a {\sqrt {b(r_1-1)}}, a {\sqrt {b(r_1-1)}}, a^2 (r_1-1) ),\\
{\bf {\rm F_5}} = ((1 - a) {\sqrt {b(r_2-1)}}, (1 - a) {\sqrt {b(r_2-1)}},
(1 - a)^2 (r_2-1) ),\\
{\bf {\rm F_6}}~=~( -a {\sqrt {b(r_1-1)}}, -a {\sqrt {b(r_1-1)}},a^2 (r_1-1) ),\\
{\bf {\rm F_7}}~=~(-(1 - a) {\sqrt {b(r_2-1)}}, -(1 - a) {\sqrt {b(r_2-1)}},
(1 - a)^2 (r_2-1) ),\\
{\bf {\rm F_8}}~=~(a {\sqrt{b(r_1-1)}} - (1 - a) {\sqrt{b
(r_2-1)}}, a {\sqrt{b(r-1)}} - (1-a) {\sqrt{b(r_2-1)}}, a^2 (r_1 - 1) + 
(1-a)^2 (r_2-1) - 2a (1-a) {\sqrt (r_1-1) (r_2-1)}),\\
{\bf {\rm F_9}}=(-a {\sqrt{b(r_1-1)}} + (1-a){\sqrt{b(r_2-1)}},
-a {\sqrt{b(r-1)}} + (1-a) {\sqrt{b(r_2-1)}},
a^2 (r_1-1) + (1-a)^2 (r_2-1) - 2a (1 - a) {\sqrt {(r_1-1)(r_2-1)}})$.

Fig. 3 shows the ($r_1, r_2$)-parameter space. The stable fixed points at
different regions are indicated in the diagram. The dotted line corresponds
to the special case $r_1=r_2$ which has been considered above. 
Note that, whereas
in the general case all the fixed points are stable in some region or other,
in the special case of $r_1=r_2$, four of the fixed 
points, $viz.$, ${\bf {\rm F_4, F_5, F_6}}$ and ${\bf {\rm 
F_7}}$, are always unstable. When one of the $r$-values go over to 
the chaotic regime, while the other $r$-value remains fairly below it, 
asymptotic synchronization with the chaotic trajectory
is observed \cite{Kowalski}. The time required to ultimately synchronize
with the chaotic attractor is a function of both the $r$-parameter values. 
The synchronization is phase- synchronization rather than
state- synchronization, as the response system chaotic attractor
is a scaled replica of the driver
system attractor. The scaling factor is $a$ for synchronization with
driving system 1, and $(1-a)$, for driving system 2. When both the $r$-values
are in the chaotic regime, the ``frustrated synchronization'' situation
occurs.

\section{Simulation Results}
For conducting simulations, the parameter values chosen were $r_1 = r_2 =
28$, $\sigma = 10$ and $b = 8/3$. The trace of the Jacobian (which
is equal to the sum of the Lyapunov exponents) for the total system,
including the driver and response systems, is -40.0. So the
overall system is diffusive and possesses an attractor. The competition
parameter $a$ was varied in the interval $[0,1]$. The differential equations
were numerically solved using the fourth-order Runge-Kutta method with 
step-size = 0.025. The phase-space trajectory
of the responding system $(x_3,y_3,z_3)$
was observed with different values of $a$ from $t$ = 0 to $t$ = 100 . At the
limit $a$ = 0 (or 1) the responding system trajectory is identical to that
of a unperturbed Lorenz system (fig. 1(a)). However, as $a \rightarrow 0.5$
(where maximal competition occurs), the trajectory deviates more and more
from the standard Lorenz form. At $a$ = 0.5, the trajectory moves in
a complicated path around the fixed points ${\bf {\rm F_1}}$,
${\bf {\rm F_2}}$ and ${\bf {\rm F_3}}$ (note that, at $a$ = 0.5, ${\bf
{\rm F_8 = F_9 = F_1}}$ (fig. 1(b)).
It appears that for $a$=0.5, the z-variable time-series is much more
correlated.
This becomes clearer on taking a Fourier transform of the data. The power
spectral density of the frustrated attractor time-series is low in the
high-frequency end compared to the unperturbed system time-series.

The Lyapunov exponents were calculated using Gram-Schmidt technique to create
an orthonormal basis every 0.5 seconds of simulation time (this time interval
being roughly half the ``period'' of the Lorenz system) and then averaging
over 100 iterations. As expected, of the eight
exponents, six correspond to those for the two unperturbed driving Lorenz
systems (0.84, 0, -14.51). The remaining two exponents are the conditional
Lyapunov exponents of the responding system : -8/3 and -10. This implies
the robustness of the ``frustrated'' attractor - as any deviation from
the attractor rapidly diminishes.

To study the degree of synchronization, $z$-coordinates of the responding
system state ($z_3$)were plotted against the $z$-coordinates of each of the
driver system states ($z_1,z_2$), for different values of $a$. 
If the two are
synchronized, the plot gives a straight line. This suggests that the linear
correlation coefficients, $r$,
between the driver and response system time series, can be used
to obtain a quantitative measure of synchronization.
The linear correlation coefficient between two time series 
data $x(t)$ and $y(t) (t=1, {\ldots}, n)$, is given by
$$
r_{x,y} = {\frac {1}{n}} {\frac {\sum_{i=1}^n (
x(i) - \bar{x}) (y(i) - \bar{y})}{\sigma_x \sigma_y}},
$$
where $\bar{x}$ and $\sigma_x$ are the mean and standard deviation
respectively, for the time series $x(t)$. 
A measure of desynchronization is defined as 
\beq
\delta = 1 - r_{z_2,z_3}.
\eeq
At $a$=0, where there is exact synchronization between driver system 2 and
the response system, $\delta = 0$. This is a particularly robust measure,
as $\delta \rightarrow 0$ for both state- and phase- synchronization.
The variation of $\delta$ with $a$
is shown in a logarithmic plot (fig. 4). The linear nature of the curve
over at least 3 orders of magnitude as $a \rightarrow 0$,
indicates the presence of a power-law scaling relation of the form:
\beq
\delta \sim a^{\beta},
\eeq
where the scaling exponent, $\beta$ $\simeq$ 2.0. The scaling exponent
was also obtained for $r$= 50 and 70. In both cases, $\beta$ $\simeq$ 2.0
within simulation error. 

Another interesting feature studied was the fractal correlation dimension
of the frustrated attractor (fig. 5), calculated using the FD3 (ver. 0.3)
software \cite{fd3}. For the unperturbed Lorenz system,
this is very close to 2, as the attractor is almost 2-dimensional.
As $a$ increases from 0 to 0.5, the attractor deviates from this
two-dimensional shape, which can be quantitatively measured by
the correlation dimension. As $a \rightarrow 0.5$, the attractor
structure stretches out more and more over the three-dimensional space.
This type of enhanced diffusion in phase space seems to be a generic
feature of frustration in chaotic systems,
and has been reported previously in the case of Coupled
Map Lattices \cite{Bersini}.

The simulations also showed the robustness of the ``frustrated'' attractor.
Starting from different initial conditions, the response system
trajectory was found to converge to the
same attractor structure. This indicates that even in the absence of
exact synchronization with any of the driver systems, the response
system trajectory can be used as a chaotic masking signal for secure
communication \cite{Cuomo}. 
This was verified by adding a small amplitude periodic
signal (e.g., a sine wave of frequency $\omega = 1/200$) 
to the response system
$y$-variable time series. The resultant time series appears to be devoid
of any periodic component (fig. 6, top).
It is then used to drive another Lorenz system,
and the $x$-variable time series of the two systems are subtracted
from each other to retrieve the original signal (fig. 6, bottom).
The modulation of the competition parameter,$a$, by a binary signal
for chaotic switching, is another possibility of using the
competitive scheme for secure communication.

\section{Discussion}
The competitive scheme described here for $y$-variable coupling was also
implemented for $x$- and $z$-coupling of Lorenz systems. In the former,
similar generalized attractor structure was observed, while in the latter,
where the Pecora-Carroll synchronization does not work, no such structure
could be observed.
The work done here on coupled Lorenz systems can be extended to other
systems defined by autonomous set of differential equations as well as
discrete maps. However, it might be interesting to consider the result of
competition in synchronizing non-autonomous systems (e.g., the Duffing
oscillator). As such systems already have a forcing term present,
which brings about the onset of chaos, the introduction of additional
forcing terms can lead to qualitatively new behavior.

Competitive synchronization in extended systems might also lead to
interesting
phenomena. Lattices of (globally or diffusively) coupled chaotic elements,
where each element can be used both to drive other elements, as well as
respond to driving signals from yet another set of elements, and hence by a
series of feedbacks drive its own driving systems, will serve to 
illustrate interactions between multiple competing synchronizing feedback
loops. The motivation for such a study is that, in the human brain,
synchronization of activity among different neurons  appear to have an
important functional role in the proper performance of perceptual tasks.
It is to be noted that, single neurons are capable of chaotic behavior.
As the brain is composed of densely connected networks of neurons,
there is bound to be competitive synchronizing interactions between
neural assemblies \cite{SinhaKar}. A dynamic competition parameter,
which causes synchronization-desynchronization transitions between various
neural sub-assemblies, is a possible mechanism for information
processing in biological systems. The resultant dynamics will be
radically different
from the one we are led to expect by observing the dynamics of single
neurons or small groups of neurons.

The above work describes the simplest competitive scenario which can
show a qualitatively different dynamics from that in the non-competitive
situation. It is at present not known how the nature of synchronization
and the attractor structure of the responding system might be
altered by increasing the number of competing driver systems.
In the brain, where each neuron is connected to $\sim ~{10}^4$
other neurons, the competitive situation is bound to be far more
complicated. The manner in which such an extremely competitive
synchronization scenario might influence the way in which neural
networks perform computations and process information is a very
interesting problem for the future.
 
\vs{1.5cm}

{\large{\bf Acknowledgments}}\\

I would like to acknowledge the helpful comments and suggestions of 
Mr. S. Kar (East India Pharmaceuticals, Calcutta) and Prof.
J. K. Bhattacharya (IACS, Calcutta). I would also like to thank
Prof. S. K. Pal (MIU, ISI) for his constant encouragement.

\vs{2cm}
{\Large {\bf List of Figures}}\\

{\bf Fig. 1} The response system attractor for (a) $a$ = 1.0 and
(b) $a$ = 0.5 ($r_1 = r_2 = 28, \sigma = 10, b =8/3$).

{\bf Fig. 2} The $z$-coordinate of fixed points of 
the response system for $0 \leq a \leq 1$.

{\bf Fig. 3} The $(r_1, r_2)$-parameter space showing the stable fixed
points of the response system at different regions.

{\bf Fig. 4} Log-scale plot of desynchronization ($\delta$) for $0<a \leq 1$.
The power-law scaling relation (with characteristic exponent, $\beta \sim$
2.0) is indicated by the solid line fitted to the simulation data.

{\bf Fig. 5} Correlation dimension of the response system attractor for
$0 \leq a \leq 1$.

{\bf Fig. 6} Chaotic masking: the $x$-variable time series of response
system (top); the periodic signal obtained by subtracting the regenerated
time series from the chaotic carrier wave (bottom).

\ed
\begin{thebibliography}{99}
\bibitem{Pecora90} L. M. Pecora and T. L. Carroll, {\it Phys.
Rev. Lett.} {\bf 64} (1990) 821.
\bibitem{Fujisaka} H. Fujisaka and T. Yamada, {\it Prog.
Theo. Phys.} {\bf 69} (1983) 32.
\bibitem{Pecora91} L. M. Pecora and T. L. Carroll, {\it Phys.
Rev. A} {\bf 44} (1991) 2374.
\bibitem{HeVaidya} R. He and P. G. Vaidya, {\it Phys. 
Rev. A} {\bf 46} (1992) 7387.
\bibitem{Cuomo} K. M. Cuomo, A. V. Oppenheim and S. H. Strogatz, {\it
IEEE Trans. on Circuits \& Systems II} {\bf 40} (1993) 626.
\bibitem{HuQuHe} G. Hu, Z. Qu and K. He, {\it Int. J. Bif.
Chaos} {\bf 5} (1995) 901.
\bibitem{Lorenz} E. N. Lorenz, {\it J. Atmospheric
Sci.} {\bf 20} (1963) 130.
\bibitem{Sparrow} C. Sparrow, {\it The Lorenz Equations:
Bifurcations, Chaos, and Strange Attractors}
(Springer, New York, 1982).
\bibitem{Kowalski} J. M. Kowalski, G. L. Albert and G. W. Gross,  
{\it Phys. Rev. A} {\bf 42} (1990) 6260.
\bibitem{fd3} ftp://ftp.immt.pwr.wroc.pl/pub/fractal.
\bibitem{Bersini} H. Bersini and V. Calenbuhr, {\it J. Theo. Biol.}
{\bf 188} (1997) 187.
\bibitem{SinhaKar} S. Sinha and S. Kar, in: {\it Methodologies
for the conception, design and application of intelligent systems},
eds. T. Yamakawa and G. Matsumoto 
(World Scientific, Singapore, 1996) p. 700.
\end{thebibliography}
